\documentclass{aa}
\usepackage{graphicx}
\usepackage{times}
\begin{document}

\def\3c{3C\,273}
\def\nh{$N\dmrm{H}$}
\def\Lya{Ly$\alpha$}
\def\Hb{H$\beta$}
\def\Hg{H$\gamma$}
\def\feii{Fe\,{\sc ii}}
\def\civl{C\,{\sc iv} $\lambda$1549}
\def\heii{He\,{\sc ii}}
\def\heiil{He\,{\sc ii} $\lambda$4686}
\def\mgii{Mg\,{\sc ii}}
\def\mgiil{Mg\,{\sc ii} $\lambda$2798}
\def\Lrest{$\lambda\dmrm{rest}$}
\def\Lobs{$\lambda\dmrm{obs}$}
\def\kms{\mbox{km\,s$^{-1}$}}
\def\Hubble{\mbox{km\,s$^{-1}$\,Mpc$^{-1}$}}
\def\ergs{\mbox{erg\,s$^{-1}$}}
\def\ergcms{\mbox{erg\,cm$^{-2}$\,s$^{-1}$}}
\def\ergcmsA{\mbox{erg\,cm$^{-2}$\,s$^{-1}$\,\AA$^{-1}$}}
\def\ergcmsHz{\mbox{erg\,cm$^{-2}$\,s$^{-1}$\,Hz$^{-1}$}}
\def\phcmskeV{\mbox{photons\,cm$^{-2}$\,s$^{-1}$\,keV$^{-1}$}}
\def\ltsim{\raisebox{-.5ex}{$\;\stackrel{<}{\sim}\;$}}
\def\gtsim{\raisebox{-.5ex}{$\;\stackrel{>}{\sim}\;$}}
\newcommand{\mrm}[1]{\ifmmode \mathrm{#1} \else $\mathrm{#1}$\fi}
\newcommand{\dmrm}[1]{_{\mathrm{\,#1}}}
\newcommand{\umrm}[1]{^{\mathrm{\,#1}}}

\thesaurus{20(04.01.1; 11.01.2; 11.17.4 3C 273; 13.18.1; 13.21.1; 13.25.2)}

\title{30 years of multi-wavelength observations of 3C\,273\thanks{Data available at: http://obswww.unige.ch/3c273/}}
\author{
M. T\"urler \inst{1,2} \and
S. Paltani \inst{1,2} \and
T.J.-L. Courvoisier \inst{1,2} \and
M.F. Aller \inst{3} \and
H.D. Aller \inst{3} \and
A. Blecha \inst{1} \and
P. Bouchet \inst{4,5} \and
M. Lainela \inst{6} \and
I.M. McHardy \inst{7} \and
E.I. Robson \inst{8,9} \and
J.A. Stevens \inst{9,10} \and
H. Ter\"asranta \inst{11} \and
M. Tornikoski \inst{11} \and
M.-H. Ulrich \inst{12} \and
E.B. Waltman \inst{13} \and
W. Wamsteker \inst{14} \and
M.C.H. Wright \inst{15}
}
\institute{
Geneva Observatory, ch. des Maillettes 51, CH-1290 Sauverny, Switzerland \and
\textit{INTEGRAL} Science Data Centre, ch. d'\'Ecogia 16, CH-1290 Versoix, Switzerland \and
University of Michigan, Department of Astronomy, 817 Dennison Building, Ann Arbor, MI 48\,109, USA \and 
European Southern Observatory, Casilla 19\,001, Santiago 19, Chile \and 
Cerro Tololo Inter-American Observatory, Casilla 603 -- La Serena, Chile \and 
Tuorla Observatory, V\"ais\"al\"antie 20, FIN-21\,500 Piikki\"o, Finland \and 
Department of Physics, University of Southampton, Southampton SO9 5NH, United Kingdom \and 
Centre for Astrophysics, University of Central Lancashire, Preston, PR1 2HE, United Kingdom \and 
Joint Astronomy Centre, 660 North A`oh$\bar{\mbox{o}}$k$\bar{\mbox{u}}$ Place, University Park, Hilo, Hawaii 96\,720, USA \and 
Mullard Space Science Laboratory, University College London, Holmbury St. Mary, Dorking, Surrey RH5 6NT, United Kingdom \and 
Mets\"ahovi Radio Observatory, Mets\"ahovintie, FIN-02\,540 Kylm\"al\"a, Finland \and 
European Southern Observatory, Karl-Schwarzschild-Strasse 2, D-85\,748 Garching bei M\"unchen, Germany \and 
Remote Sensing Division, Naval Research Laboratory, Washington, DC 20\,375-5351, USA \and 
ESA-Vilspa, P.O. Box 50\,727, E-28\,080 Madrid, Spain \and 
Department of Astronomy, University of California, Berkeley, CA 94\,720, USA 
}
\offprints{M. T\"urler (ISDC)}
\mail{Marc.Turler@obs.unige.ch}
\date{Received date / Accepted date}
\maketitle

\begin{abstract}
We present a wide multi-wavelength database of most observations of the quasar \3c\ obtained during the last 30 years.
This database is the most complete set of observations available for an active galactic nucleus (AGN).
It contains nearly 20\,000 observations grouped together into 70 light curves covering 16 orders of magnitude in frequency from the radio to the $\gamma$-ray domain.

The database is constituted of many previously unpublished observations and of most publicly available data gathered in the literature and on the World Wide Web (WWW).
It is complete to the best of our knowledge, except in the optical (UBV) domain where we chose not to add all observations from the literature.
In addition to the photometric data, we present the spectra of \3c\ obtained by the International Ultraviolet Explorer (IUE) satellite.
In the X-ray domain, we used the spectral fit parameters from the literature to construct the light curves.

Apart from describing the data, we show the most representative light curves and the average spectrum of \3c.
The database is available on the WWW in a homogeneous and clear form and we wish to update it regularly by adding new observations.

\end{abstract}

\begin{figure*}[tb]
\includegraphics[width=\hsize]{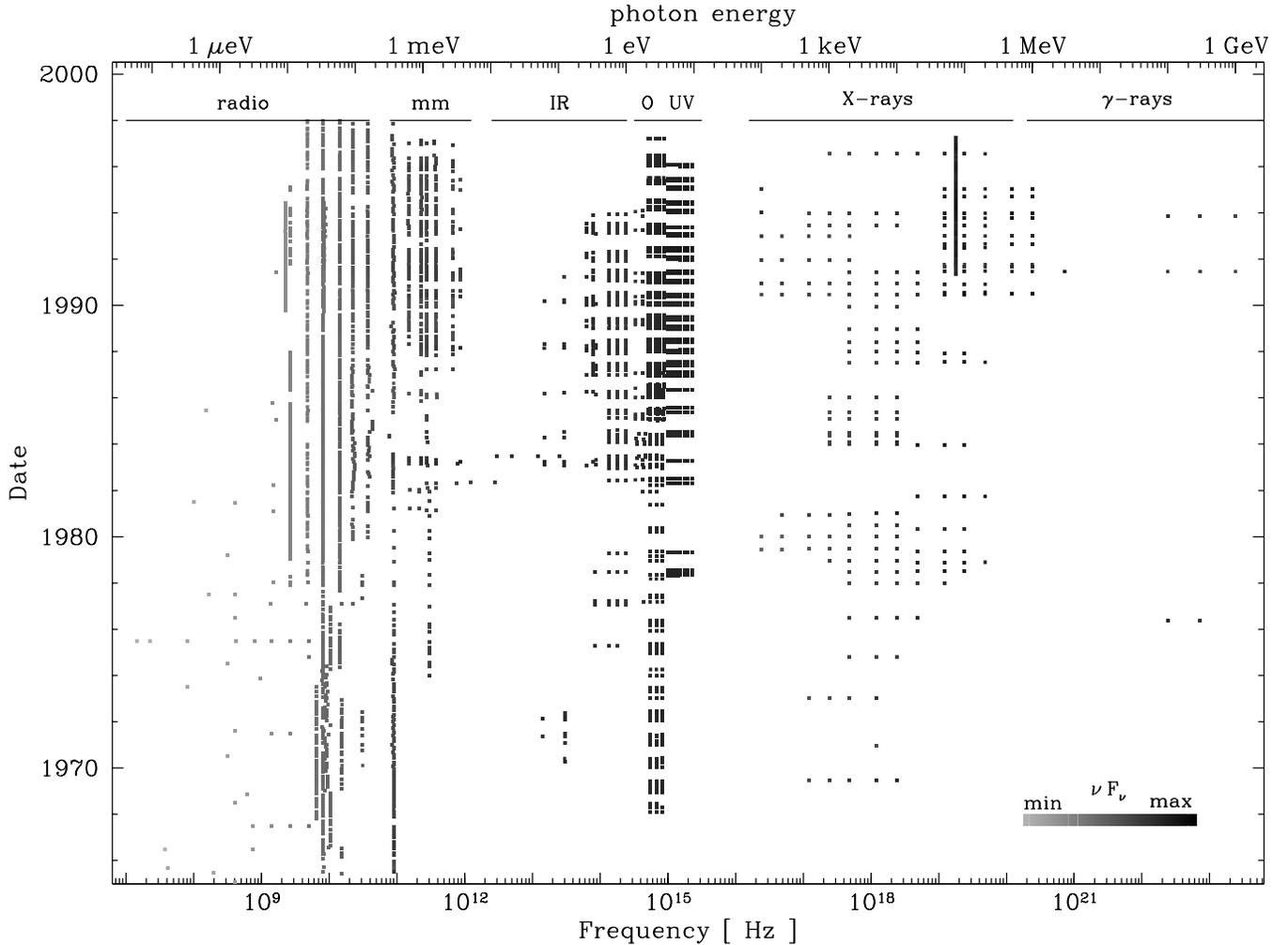}
\caption{Time versus frequency distribution of all observations of \3c\ presently in the database. This figure summarizes more than 30\,years of observations in the spectral range from 10\,MHz to 1\,GeV, covering 16 orders of magnitude in frequency. The relative intensity of the observed flux (log($\nu\,F_{\nu}$)) is given by a greyscale}
\label{allobs}
\end{figure*}

\section{\label{intro}Introduction}

\object{3C\,273} is the brightest quasar on the sky, with a mean V band magnitude of 12.9.
It can be easily observed from both hemispheres, thanks to its position very close to the celestial equator: $\alpha$\,=\,12$^{h}$29$^{m}$06.70$^{s}$, $\delta$\,=\,$+$02$\degr$03$\arcmin$08.6$\arcsec$ (J2000.0) and at high galactic latitude: $l$\,=\,289.95, $b$\,=\,$+$64.36.
The redshift of \3c\ is $z$\,=\,0.158, which corresponds to an effective distance of 440\,$h^{-1}$\,Mpc, with a Hubble constant defined as $H_0=100\,h\,\Hubble$ and a deceleration parameter of $q_0$\,=\,0.05.
The mean bolometric flux received from \3c\ between 10$^{7}$ and 10$^{25}$\,Hz (see Fig.~\ref{avspect}) is 1.9\,10$^{-9}$\,\ergcms.
If the whole spectrum of \3c\ is emitted isotropically, this flux would correspond to a bolometric luminosity of 6.0\,10$^{46}\,h^{-2}$\,\ergs.
The host galaxy of \3c\ is an elliptical (E4) galaxy, which has an outer radius of about 15$\arcsec$ and a V band magnitude of 16.4 (Bahcall et al. \cite{BKS97}).

\3c\ has a jet with apparent superluminal motion that ends with a hot spot called \object{3C\,273A}.
It is among the few active galactic nuclei (AGN) detected at energies above 100\,MeV by the Energetic Gamma-Ray Experiment Telescope (EGRET) on board the Compton Gamma-Ray Observatory (CGRO) and is thus often classified as a blazar despite its prominent blue-bump and its strong emission-lines.
This coexistence of blazar- and Seyfert-like properties makes \3c\ be a very interesting but complex object.

The detection of strong ultraviolet variability in 1982 (Courvoisier \& Ulrich \cite{CU85}) was at the origin of the multi-wavelength monitoring campaign that started in 1983.
This observation effort, which is still going on, led to several publications on the multi-wavelength properties of \3c\ (Courvoisier et al. \cite{CTR87}, \cite{CRB90}; Lichti et al. \cite{LBC95}; von Montigny et al. \cite{VAA97}).
More detailed variability studies in specific spectral domains were also possible with these observations (e.g. Turner et al. \cite{TWC90}; Cappi et al. \cite{CMO98} (X-rays); Ulrich et al. \cite{UCW88} (ultraviolet); Courvoisier et al. \cite{CRB88} (optical and infrared); Robson et al. \cite{RLG93} (infrared to radio); Stevens et al. \cite{SRG98} (millimetre to radio)).

The end of the IUE operations and the writing of a review paper on \3c\ (Courvoisier \cite{C98}) is a good opportunity to make all these data available.
The aim of this contribution is to present and to maintain a high quality publicly available database of most observations of \3c.
Publishing such a wide database is not a goal in itself, but has the purpose to stimulate variability analyses by a large community of astronomers.
The huge effort of observing \3c\ during more than 30 years should lead to a better understanding of this object and hence of AGN in general.
The detailed studies of the blue-bump variability (Paltani et al. \cite{PCW98}) and of the millimetre-to-radio flaring behaviour (T\"urler et al. in preparation) are examples of what can be done with these data and we hope that this analysis effort will continue elsewhere, thanks to this database.

\section{\label{R}Radio observations}

\begin{figure*}[tb]
\includegraphics[width=\hsize]{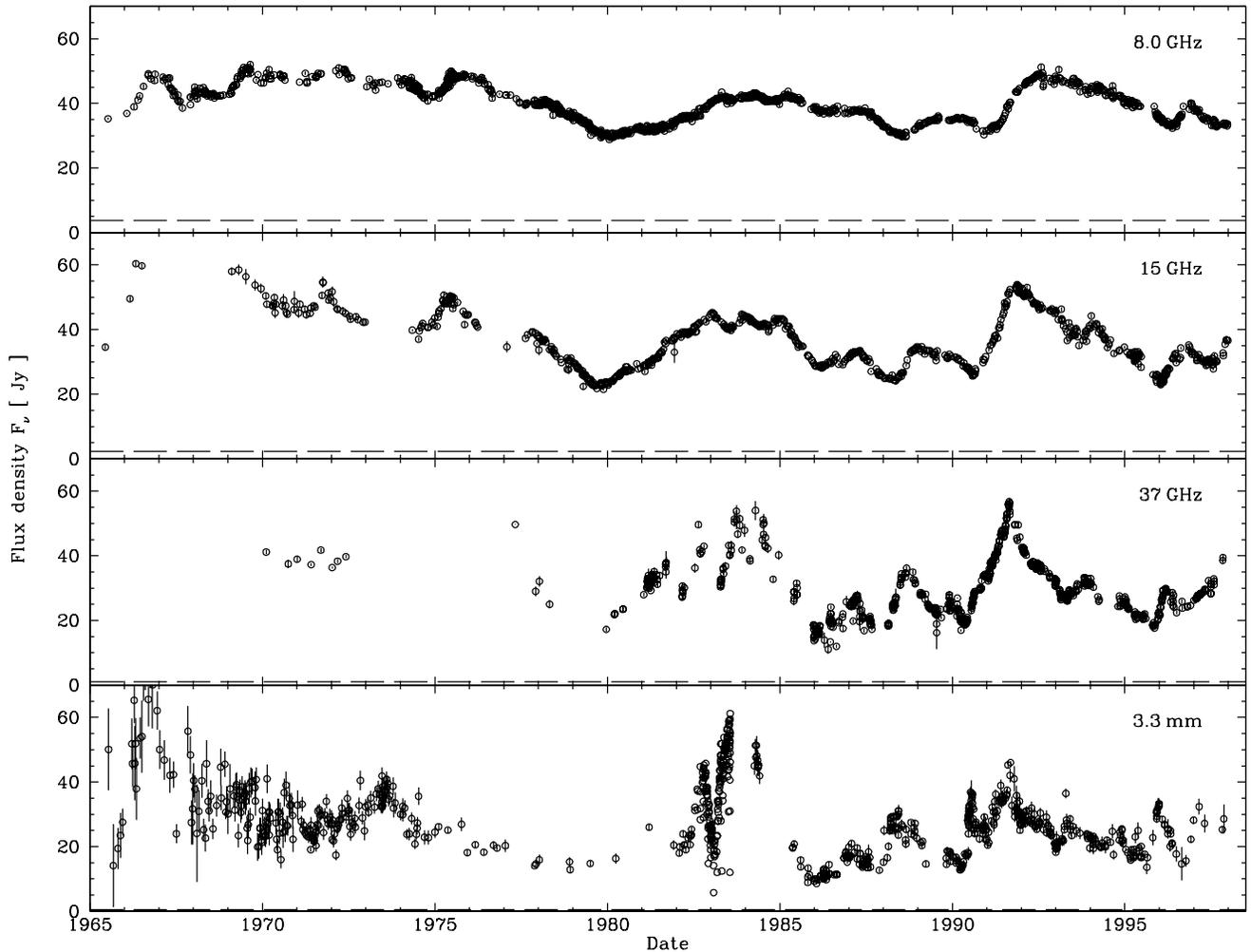}
\caption{Four characteristic light curves of the radio-to-millimetre behaviour of \3c\ shown on a same scale from 1965 to 1998.
The dashed line is the contribution from the jet (3C\,273A) (see Sect.~\ref{R})}
\label{lcrmm}
\end{figure*}

We grouped together all radio measurements into 17 light curves from 15\,MHz to 37\,GHz (see Table \ref{tabrmm}).
The 8.0\,GHz light curve obtained with the 26\,m radio telescope of the University of Michigan Radio Astronomy Observatory (UMRAO) is the most complete light curve of \3c\ obtained by a single observatory during more than 30\,years (see Fig. \ref{lcrmm}).
It is therefore given alone in the database, whereas other observations around 8\,GHz are included in the 10\,GHz light curve.
The UMRAO monitoring of \3c\ at 14.5\,GHz and 4.8\,GHz started respectively in 1974 and in 1978.
The 5\,GHz and the 15\,GHz light curves contain mainly these observations.
Details on the instrumentation and the calibration used at the UMRAO are given by Aller et al. (\cite{AAL85}) together with the data obtained until 1984.
We do not include here the polarization observations, which are publicly available at the UMRAO Database Interface on the WWW at http://www.astro.lsa.umich.edu/obs/radiotel/umrao.html.

Shorter wavelengths radio observations at 22.2 and 36.8\,GHz were performed since 1980 both with the 14\,m telescope of the Mets\"ahovi Radio Observatory, Finland and with the 22\,m telescope of the Crimean Astrophysical Observatory, Ukraine.
The 22\,GHz and 37\,GHz light curves are very well sampled since 1986 except for a gap in the summer of 1994 due to the replacement of the Mets\"ahovi antenna (see Fig. \ref{lcrmm}).
The observations during 1980--85 and during 1985--90 are respectively published in Salonen et al. (\cite{STU87}) and in Ter\"asranta et al. (\cite{TTV92}), together with details on the measurement methods and the calibrations.
Calibrations at 22\,GHz were usually performed with the nearby source 3C\,274 (Virgo\,A, M\,87), whose flux was taken to be 21.7\,Jy.
Since there might have been a recent outburst in 3C\,274, the 22\,GHz data from Mets\"ahovi presented here are only calibrated with the primary calibrator DR\,21.
We therefore have the same calibration procedure at 22 and 37\,GHz.
The differences between the two calibrations are generally within the uncertainties.

Daily observations of \3c\ were performed by the Green Bank Interferometer (GBI) at 2.7\,GHz and at 8.1\,GHz from 1979 to 1988.
This huge data set was published by Waltman et al. (\cite{WFJ91}).
Additional observations carried out with new receivers at 2.25\,GHz and at 8.3\,GHz from 1989 to 1994 are also included in the database.
The GBI light curves display brightness dips, which occur when the sun is too close to \3c\ on the sky.
Since the GBI is an interferometer, it does not measure the total flux of an extended source like \3c.
It is therefore difficult to compare the GBI measurements with single dish telescope observations.
For clarity, the GBI data are stored in separate files.

Other repeated radio observations from the literature were added to the database.
Observations at 2.7, 4.75 and 10.55\,GHz from the 100\,m telescope at Effelsberg, Germany reported in von Montigny et al. (\cite{VAA97}) were added to the 2.5\,GHz, the 5\,GHz and the 10\,GHz light curves respectively.
The 5\,GHz and 10\,GHz light curves also contain earlier observations at 6.6\,GHz and at 10.6\,GHz from the 46\,m telescope of the Algonquin Radio Observatory (Medd et al. \cite{MAH72}; Andrew et al. \cite{AMH78}).
Observations at 7.8, 7.9 and 15.5\,GHz from the 37\,m antenna of the Haystack Radio Observatory are included in the 10\,GHz and in the 15\,GHz light curves (Allen \& Barrett \cite{AB66}; Dent \& Kapitzky \cite{DK76}; Dent \& Kojoian \cite{DK72}; Dent et al. \cite{DKK74}).
We also added to the 22\,GHz and 37\,GHz light curves the 22 and 44\,GHz observations from the 13.7\,m Itapetinga radio telescope, Brazil (Botti \& Abraham \cite{BA88}) and the 24\,GHz observations from the UMRAO (Haddock et al. \cite{HAA87}).
Finally, we included in the 37\,GHz light curve a few earlier observations at 31.4\,GHz made with the 11\,m antenna of the National Radio Astronomical Observatory (NRAO) at Kitt Peak (Dent \& Hobbs \cite{DH73}).

We did not include all the isolated observations from early radio catalogues.
We added only the flux measurements reported by K\"uhr et al. (\cite{KWP81}), since they are all recalibrated to the scale of Baars et al. (\cite{BGP77}), and the total flux densities (core plus jet) reported by Conway et al. (\cite{CGP93}).
At very low frequency ($<$\,30\,MHz), we added the observations from Braude et al. (\cite{BMS79}), but multiplied by the scaling factor of 1.23 used for other objects by K\"uhr et al. (\cite{KWP81}).
In the 100--1000\,MHz range other isolated observations are from Artyukh (\cite{A84}), Dennison et al. (\cite{DBL81}), Fanti et al. (\cite{FFM79}, \cite{FFF81}), Fisher \& Erickson (\cite{FE80}) and Hunstead (\cite{H72}).
Above 1\,GHz, some isolated radio observations were found in Jones et al. (\cite{JRO81}), Landau et al. (\cite{LJE83}) and Lichti et al. (\cite{LBC95}).
All these data were included in the respective light curves.

The radio light curves of \3c\ at 8.0\,GHz, 15\,GHz and 37\,GHz are shown in Fig.~\ref{lcrmm}.
The contribution from the jet (3C\,273A) was derived from Fig.~A1 of Conway et al. (\cite{CGP93}).
It is a broken power law with a spectral index $\alpha\dmrm{jet}$ of 0.67 below 735\,MHz (29.0\,Jy) and of 0.85 at higher frequencies.
The flux density of 3C\,273A declines strongly in the infrared-to-optical domain with $\alpha\dmrm{jet}$\,$\sim$\,4, as shown in Fig.~4f of Meisenheimer et al. (\cite{MRH89}).
Fig.~\ref{avspect} shows that the jet component is dominant below $\sim$\,1\,GHz, whereas it becomes negligible ($<$\,5\,\% of $\overline{F_{\nu}}$) above 22\,GHz.

\begin{table}[tb]
\caption{The radio and mm/submm light curves of \3c\ in the database. The light curves are characterized by the date range (Epoch) between the first and the last observation, the number $N$ of observations, the mean frequency $\overline{\nu}$ in Hz, the mean flux density $\overline{F_{\nu}}$ in Jy and the dispersion $\sigma_{\nu}$ in Jy}
\label{tabrmm}
\begin{flushleft}
\begin{tabular}{@{}lcrcrr@{}}
\hline
\rule[-0.7em]{0pt}{2.0em}Light curve& Epoch& \multicolumn{1}{c}{$N$}& $\overline{\nu}$& \multicolumn{1}{c}{$\overline{F_{\nu}}$}& \multicolumn{1}{c}{$\sigma_{\nu}$}\\
\hline
\rule{0pt}{1.2em}15\,MHz& 1975--75& 3& 1.47\,10$^{7}$& 451.4& 57.5\\
22\,MHz& 1975--75& 2& 2.25\,10$^{7}$& 316.1& 33.1\\
40\,MHz& 1965--66& 2& 3.98\,10$^{7}$& 204.7& 69.7\\
100\,MHz& 1962--81& 4& 9.95\,10$^{7}$& 138.2& 33.6\\
200\,MHz& 1963--85& 8& 1.95\,10$^{8}$& 81.9& 9.3\\
400\,MHz& 1962--81& 16& 3.97\,10$^{8}$& 59.8& 4.4\\
800\,MHz& 1963--75& 11& 7.58\,10$^{8}$& 48.9& 3.2\\
1.5\,GHz& 1962--91& 11& 1.48\,10$^{9}$& 40.8& 5.5\\
GBI\,2\,GHz& 1979--94& 2831& 2.59\,10$^{9}$& 35.0& 5.7\\
2.7\,GHz& 1964--95& 37& 2.70\,10$^{9}$& 41.5& 3.3\\
5\,GHz& 1967--97& 548& 5.00\,10$^{9}$& 38.4& 4.3\\
8.0\,GHz& 1965--97& 1242& 8.00\,10$^{9}$& 39.3& 5.6\\
GBI\,8\,GHz& 1979--94& 3061& 8.14\,10$^{9}$& 31.0& 3.9\\
10\,GHz& 1965--94& 218& 9.57\,10$^{9}$& 48.0& 4.5\\
15\,GHz& 1965--97& 776& 1.46\,10$^{10}$& 36.2& 8.3\\
22\,GHz& 1977--97& 773& 2.23\,10$^{10}$& 33.5& 8.3\\
37\,GHz& 1970--97& 774& 3.69\,10$^{10}$& 29.1& 8.6\\
3.3\,mm& 1965--97& 875& 8.91\,10$^{10}$& 28.6& 11.5\\
2\,mm& 1981--97& 203& 1.50\,10$^{11}$& 17.9& 6.3\\
1.3\,mm& 1981--97& 375& 2.30\,10$^{11}$& 13.2& 5.5\\
1.1\,mm& 1973--97& 316& 2.76\,10$^{11}$& 13.5& 6.4\\
0.8\,mm& 1981--97& 243& 3.74\,10$^{11}$& 10.3& 4.9\\
0.45\,mm& 1987--96& 72& 6.66\,10$^{11}$& 5.4& 2.5\\
0.35\,mm& 1982--95& 23& 8.34\,10$^{11}$& 9.6& 6.7\\
\hline
\end{tabular}
\end{flushleft}
\end{table}

\section{\label{MM}Millimetre and submillimetre observations}

All the millimetre/submillimetre (mm/submm) observations of \3c\ were grouped together into seven light curves: 3.3\,mm (90\,GHz), 2.0\,mm (150\,GHz), 1.3\,mm (230\,GHz), 1.1\,mm, 0.8\,mm, 0.45\,mm and 0.35\,mm (see Table \ref{tabrmm}).
The 3.3\,mm light curve is shown in Fig. \ref{lcrmm}.

All the submillimetre observations, as well as most observations at 1.1\,mm, 1.3\,mm and 2.0\,mm were performed on Mauna Kea, Hawaii.
The QMC/Oregon photometer (Ade et al. \cite{AGC84}) was used on the 3.8\,m United Kingdom Infrared Telescope (UKIRT) until 1985.
Since January 1986, a new common user photometer UKT14 (Duncan et al. \cite{DSR90}) was installed on the UKIRT, before being moved in March 1988 to the 15\,m James Clerk Maxwell Telescope (JCMT).
Finally, in July 1996, UKT14 was replaced on the JCMT by the Submillimetre Common-User Bolometer Array (SCUBA), described by Robson et al. (\cite{R98}).
Details on the observations performed with the UKT14 photometer both on the JCMT and the UKIRT, as well as calibration techniques are given by Robson et al. (\cite{RLG93}).
The earlier observation techniques with the QMC/Oregon photometer are described in Robson et al. (\cite{RGC83}).

A second important source of millimetre observations at 90 and 230\,GHz is the 15\,m Swedish-ESO Submillimetre Telescope (SEST) on the European Southern Observatory (ESO) site at La Silla, Chile.
Until June 1995, the 90\,GHz observations were obtained using a dual polarization Schottky receiver, and since then measurements were made with a superconductor-insulator-superconductor (SIS) receiver.
As a back-end, wide band (1\,GHz) acousto-optic spectrometers (AOS) were used.
For the 230\,GHz observations a Schottky receiver and a wide band AOS were initially used, but since 1991 measurements are mainly obtained with a single channel bolometer.
The flux density measurements were made in a dual beam-switching mode, and calibrated against planets.
All the SEST observations until June 1994 were published by Tornikoski et al. (\cite{TVT96}), together with more details on the observation techniques.

Interferometer observations of \3c\ at around 90\,GHz (75--115\,GHz) were obtained since 1986 with the Berkeley-Illinois-Maryland Association (BIMA) millimetre array located at Hat Creek, California (Welch et al. \cite{WTP96}). 
The \3c\ flux densities were measured relative to planets assuming the
planetary brightness temperatures given by Ulich (\cite{U81}).
The larger error bars include the uncertainties in the correction for atmospheric decorrelation on the longer baselines.
All BIMA observations are included in the 3.3\,mm light curve.
This light curve also contains several unpublished 87.3\,GHz observations performed since 1985 at the Mets\"ahovi Radio Observatory, Finland (Ter\"asranta et al. \cite{TTV92}).

An other major source of \3c\ data in the millimetre domain are the published observations at 90, 150 and 230\,GHz obtained using the heterodyne detectors of the ``Institut de Radio-Astronomie Millimétrique'' (IRAM) 30\,m telescope at Pico Veleta, Spain.
Details on the observations and the calibrations are given by Steppe (\cite{S92}), whereas the measurements were published by Steppe et al. (\cite{SSC88}, \cite{SLM92}, \cite{SPS93}) and by Reuter et al. (\cite{RKS97}).

Other mm/submm observations from the literature were added to the database.
In the 3.3\,mm light curve, we included the huge set of observations made at 90\,GHz with the Aerospace 4.6\,m telescope from 1965 to 1975 (Epstein et al. \cite{EFM82}) and the less frequent measurements made at 85.2 or 90\,GHz with the 11\,m NRAO telescope located on Kitt Peak (Hobbs \& Dent \cite{HD77}).
The 3.3\,mm light curve also contains the very well sampled observations performed in 1982--83 with the 25\,m telescope at Hat Creek, California (Backer \cite{B84}), the 87\,GHz observations obtained in 1981--82 at the 14\,m telescope of the Five College Radio Astronomy Observatory (FCRAO) (Barvainis \& Predmore \cite{BP84}), and the 77\,GHz observations obtained at Mets\"ahovi in 1984 (Ter\"asranta et al. \cite{TVH87}).
Early monitoring of \3c\ was also performed at 1\,mm with the 5\,m Hale telescope from 1973 to 1980 (Elias et al. \cite{EEG78}; Ennis et al. \cite{ENW82}).
Early UKIRT mm/submm observations together with NRAO measurements at various wavelengths are reported by Robson et al. (\cite{RGC83}, \cite{RGB86}) and Gear et al. (\cite{GRA84}).
Other isolated millimetre observations are reported by several authors (Chini et al. \cite{CKM84}; Clegg et al. \cite{CGA83}; Courvoisier et al. \cite{CTR87}; Geldzahler et al. \cite{GW81}; Jones et al. \cite{JRO81}; Landau et al. \cite{LER80}; Owen \& Puschell \cite{OP82}; Owen et al. \cite{OPM78}; Roellig et al. \cite{RBI86}; Sherwood et al. \cite{SKG83}).
All these data were included in the respective light curves.

\section{\label{IR}Infrared observations}

The database contains infrared (IR) observations made through the eight filters J (1.25\,$\mu$m), H (1.65\,$\mu$m), K (2.2\,$\mu$m), L (3.45\,$\mu$m), L$^{\prime}$ (3.8\,$\mu$m), M (4.8\,$\mu$m), N (10\,$\mu$m) and Q (20\,$\mu$m), as well as very few far IR observations grouped together into the 60\,$\mu$m, the 100\,$\mu$m and the 240\,$\mu$m light curves (see Table \ref{tabirouv}).

The two main sources of observations in the IR are the 3.8\,m UKIRT on Mauna Kea, Hawaii and different ESO telescopes (1\,m, 2.2\,m and 3.6\,m) at La Silla, Chile.
Since the flare of 1983, \3c\ was intensively observed both from Mauna Kea and La Silla.
The UKIRT observations of \3c\ until 1993 and the ESO observations until 1990 were published by Litchfield et al. (\cite{LRS94}).
The same data are presented here, but with the recent unpublished ESO observations from 1990 to 1993 and some earlier measurements from the literature.
Following Litchfield et al. (\cite{LRS94}), all IR magnitudes were converted to fluxes, using the standard NASA Infrared Telescope Facility (IRTF) zero-magnitude fluxes, which are in Jy (10$^{-23}$\,\ergcmsHz): 1600 (J), 1020 (H), 657 (K), 290 (L), 252 (L$^{\prime}$), 163 (M), 39.8 (N) and 10.4 (Q).
The IRTF conversion has the advantage of being intermediate between the UKIRT and the ESO conversions (cf. Courvoisier et al. \cite{CRB90}).
The use of a same conversion for all measurements enables the user to easily reconvert the fluxes into magnitudes if needed.

We included in the database the IR magnitude measurements from the literature reported by Allen (\cite{A76}), Cutri et al. (\cite{CWR85}), Elvis et al. (\cite{EWM94}), Glass (\cite{G79}), Hyland \& Allen (\cite{HA82}), Kotilainen et al. (\cite{KWB92}) (12$\arcsec$ aperture), McLeod \& Rieke (\cite{MR94}), O'Dell et al. (\cite{OPS78}), Smith et al. (\cite{SBE87}) and Takalo et al. (\cite{TKD92}).
All these magnitudes were converted into fluxes using the IRTF zero-magnitude fluxes given above.
We also added the early 10 and 21\,$\mu$m flux measurements published by Rieke \& Low (\cite{RL72}), as well as other flux densities given by Courvoisier et al. (\cite{CTR87}), Robson et al. (\cite{RGC83}, \cite{RGB86}), Roellig et al. (\cite{RBI86}) and von Montigny et al. (\cite{VAA97}).
A light curve of early IR observations of \3c\ from 1967 to 1978 is shown by Neugebauer et al. (\cite{NOB79}), but unfortunately these data could not be obtained yet.

Above 20\,$\mu$m the only IR detections of \3c\ are those performed by the Kuiper Airborne Observatory (KAO) at 107 and 240\,$\mu$m (Clegg et al. \cite{CGA83}), and by the Infrared Astronomical Satellite (IRAS) at 12, 25, 60 and 100\,$\mu$m (Neugebauer et al. \cite{NMS86}; Impey \& Neugebauer \cite{IN88}; Elvis et al. \cite{EWM94}).

The obtained H band light curve is compared to higher frequency light curves in Fig.~\ref{lcirouvx}.
The contribution from the host galaxy in the H band is 6.4\,mJy.
This value corresponds to 13--14\,\% of the mean H band flux and was obtained from the host galaxy's H magnitude of 13.0 (McLeod \& Rieke \cite{MR94}) by using the H band IRTF zero-magnitude flux (1020\,Jy).
According to Fig.~1 of McLeod \& Rieke (\cite{MR95}), the stellar contribution has been very roughly estimated to peak at \Lrest\,$\sim$\,1\,$\mu$m with a value of $\sim$\,1.4\,10$^{-11}$\,\ergcms\ for \3c\ (see Fig.~\ref{avspect}).

\begin{table}[tb]
\caption{The infrared, optical and ultraviolet light curves of \3c\ in the database. The parameters are as in Table \ref{tabrmm} except that $\overline{F_{\nu}}$ and $\sigma_{\nu}$ are expressed in mJy}
\label{tabirouv}
\begin{flushleft}
\begin{tabular}{@{}lcrcrr@{}}
\hline
\rule[-0.7em]{0pt}{2.0em}Light curve& Epoch& \multicolumn{1}{c}{$N$}& $\overline{\nu}$& \multicolumn{1}{c}{$\overline{F_{\nu}}$}& \multicolumn{1}{c}{$\sigma_{\nu}$}\\
\hline
\rule{0pt}{1.2em}240\,$\mu$m& 1982--82& 1& 1.25\,10$^{12}$& 3000.0& ---~~~\\
100\,$\mu$m& 1982--83& 4& 2.95\,10$^{12}$& 2935.8& 636.0\\
60\,$\mu$m& 1983--83& 3& 5.00\,10$^{12}$& 1966.3& 210.2\\
Q (20\,$\mu$m)& 1971--90& 17& 1.44\,10$^{13}$& 826.5& 286.0\\
N (10\,$\mu$m)& 1970--91& 38& 2.94\,10$^{13}$& 387.5& 134.8\\
M (4.8\,$\mu$m)& 1983--93& 40& 6.25\,10$^{13}$& 270.2& 118.8\\
L$^{\prime}$ (3.8\,$\mu$m)& 1983--93& 107& 7.89\,10$^{13}$& 170.3& 30.4\\
L (3.45\,$\mu$m)& 1975--93& 22& 8.67\,10$^{13}$& 160.5& 33.5\\
K (2.2\,$\mu$m)& 1975--93& 195& 1.36\,10$^{14}$& 83.8& 13.5\\
H (1.65\,$\mu$m)& 1975--93& 193& 1.82\,10$^{14}$& 47.3& 8.9\\
J (1.25\,$\mu$m)& 1977--93& 191& 2.40\,10$^{14}$& 34.4& 6.4\\
I (9000\,\AA)& 1982--94& 113& 3.37\,10$^{14}$& 28.1& 5.9\\
R (7000\,\AA)& 1977--94& 86& 4.33\,10$^{14}$& 27.4& 5.1\\
G (5798\,\AA)& 1985--97& 376& 5.17\,10$^{14}$& 31.4& 3.0\\
V (5479\,\AA)& 1968--97& 636& 5.50\,10$^{14}$& 28.6& 3.2\\
V$_1$ (5395\,\AA)& 1985--97& 376& 5.56\,10$^{14}$& 29.0& 2.9\\
B$_2$ (4466\,\AA)& 1985--97& 376& 6.71\,10$^{14}$& 26.3& 2.9\\
B (4213\,\AA)& 1968--97& 623& 7.00\,10$^{14}$& 26.5& 3.1\\
B$_1$ (4003\,\AA)& 1985--97& 376& 7.49\,10$^{14}$& 27.1& 3.0\\
U (3439\,\AA)& 1968--97& 612& 8.57\,10$^{14}$& 26.5& 3.2\\
3000\,\AA& 1978--96& 209& 9.99\,10$^{14}$& 24.8& 6.5\\
2700\,\AA& 1978--96& 209& 1.11\,10$^{15}$& 23.3& 5.7\\
2425\,\AA& 1978--96& 212& 1.24\,10$^{15}$& 20.2& 5.0\\
2100\,\AA& 1978--96& 212& 1.43\,10$^{15}$& 18.5& 4.9\\
1950\,\AA& 1978--96& 252& 1.54\,10$^{15}$& 18.5& 4.2\\
1700\,\AA& 1978--96& 252& 1.76\,10$^{15}$& 16.2& 4.0\\
1525\,\AA& 1978--96& 252& 1.97\,10$^{15}$& 15.2& 3.7\\
1300\,\AA& 1978--96& 252& 2.31\,10$^{15}$& 12.0& 3.1\\
\hline
\end{tabular}
\end{flushleft}
\end{table}

\begin{figure}[tb]
\includegraphics[width=\hsize]{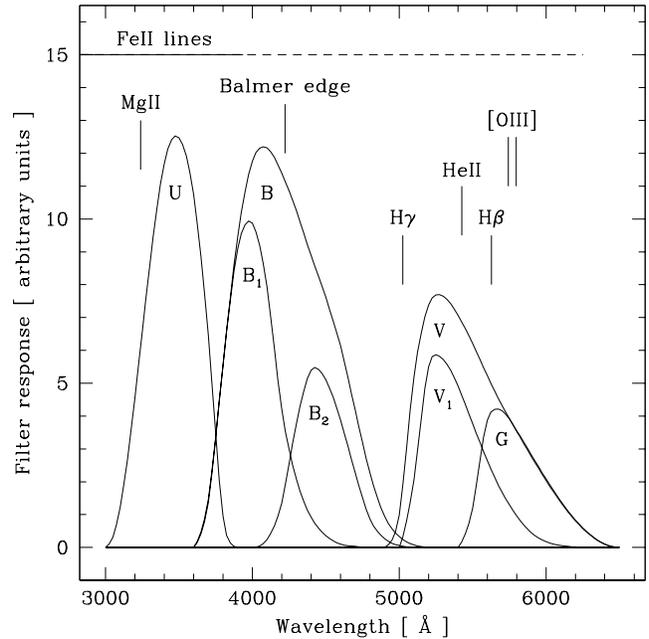}
\caption{Passband functions of the seven filters in the Geneva photometric system. The observed position of the main emission-lines in \3c\ is shown}
\label{filtres}
\end{figure}

\section{\label{O}Optical observations}

Since 1985, \3c\ is observed about once a week (when the Moon is not too bright) during its visibility season, from end of December to beginning of July with the Swiss telescope at La Silla, Chile.
Our database contains presently 376 observations made in the seven-colour Geneva photometric system from 1985 to 1997: U (3439\,\AA), B$_1$ (4003\,\AA), B (4213\,\AA), B$_2$ (4466\,\AA), V$_1$ (5395\,\AA), V (5479\,\AA) and G (5798\,\AA).
The quoted wavelength are effective wavelength calculated with the spectrum of \3c.
The stability of the Geneva photometry is extremely good, and the uncertainty on the magnitudes can be as small as 0.01, which is about 1\,\% in flux.

The response functions of the seven filters to an equiphotonic flux are shown in Fig.~\ref{filtres}.
The B and V filters are broad-band filters (width $\sim$\,300\,\AA) relatively close to the corresponding Johnson's filters.
The other filters are narrower (width $\sim$\,180\,\AA).
The U filter is comparable to Johnson's U filter, apart from the fact that its ``red'' wing is cut below the Balmer discontinuity (3647\,\AA).
B$_1$ and B$_2$ are roughly the ``blue'' and the ``red'' parts of B, and V$_1$ and G are roughly the ``blue'' and the ``red'' parts of V.

Broad-band photometry has the disadvantage of including emission-line contamination in the continuum measurements.
The observed position of the main emission-lines in \3c\ is shown in Fig.~\ref{filtres}.
Unfortunately, the flux measurements in all filters are contaminated by line emission.
Using the line parameters given by Wills et al. (\cite{WNW85}), we calculated that the \mgiil\ line contributes by 2.8\,\% in the U filter and that the broad \Hb\ line contributes respectively by 9.1\,\%, 9.7\,\% and 17.1\,\% in the V$_1$, V and G filters.
The other lines are expected to contribute much less to the V$_1$, V and G filters.
The contamination in the U, B$_1$, B and B$_2$ filters due to the \feii\ pseudo-continuum and to the Balmer lines is difficult to estimate.
Since the ultraviolet emission-lines (\Lya\ and \civl) are nearly constant in \3c\ (Ulrich et al. \cite{UCW93}; T\"urler \& Courvoisier \cite{TC98}), we do not expect significant variations of the optical lines.
Therefore, the optical flux variations are expected to be nearly pure continuum variations.

The normal photometric reduction produces magnitudes in the seven filters, which are transformed into fluxes following the calibration of Rufener \& Nicolet (\cite{RN88}) (Eq.~(6) with the $E_{\nu}$ constants of Table~8).
The uncertainties have been estimated by taking two consecutive observations in about 40\,\% of the nights; the average of the deviation in flux during these nights provided the uncertainties of the order of 1\,\%, even in the less sensitive filters.
The systematic uncertainty introduced by the absolute calibration has been estimated to be about 5 times the photometric accuracy, i.e. of the order of 5\,\% (Rufener \& Nicolet \cite{RN88}).

Other optical observations in the UBV filters from the literature were included, but only when they satisfy the following criteria.
We included only photo-electric measurements, since the photographic observations have usually much greater uncertainties (see the B band light curve of \3c\ from 1887 to 1980 shown by Angione \& Smith (\cite{AS85})).
We considered only sets of magnitude measurements made with the same instrument and having at least a few observations during the period from 1970 to 1985.
It means that we did not include isolated magnitudes, as well as sets of observations that are contemporaneous with the Geneva photometric observations, since they would only add artificial scatter to the well sampled light curves due to the use of different instruments, filter passbands and calibrations.

From all references given by Belokon' (\cite{B91}), the five following sets of magnitudes satisfy our criteria.
The two main sources of data are those of Burkhead (Burkhead \cite{B69}, \cite{B80}; Burkhead \& Hill \cite{BH75}; Burkhead \& Lee \cite{BL70}; Burkhead \& Parvey \cite{BP68}; Burkhead \& Rettig \cite{BR72}; Burkhead \& Stein \cite{BS71}) and the Crimean observations of Lyutyi (Lyutyi \cite{L76}; Lyutyi \& Metlova \cite{LM87}), which cover respectively the periods from 1968 to 1979 and from 1971 to 1986.
The uncertainty of each measurement in these two data sets was assumed to be either 0.03 or 0.05 magnitude depending on the remarks in Burkhead's papers and the presence of a ``:'' sign in Lyutyi's papers.
The other sets of magnitudes we considered are those from Mount Lemmon (Cutri et al. \cite{CWR85}; Elvis et al. \cite{EWM94}; O'Dell et al. \cite{OPS78}; Smith et al. \cite{SBE87}), from Turku (Sillanp\"a\"a et al. \cite{SHK88}), and the magnitudes from Las Campanas (Impey et al. \cite{IMT89}) with uncertainties not exceeding 0.1 magnitude.

\begin{table}[tb]
\caption{List of the magnitude corrections applied to the UBV sets of magnitudes. The number $N$ of nearly simultaneous ($\Delta\,t\leq 5$\,days) observation pairs we used is given in parentheses}
\label{UBVcorr}
\begin{tabular}{@{}lc@{ }cc@{ }cc@{ }c@{}}
\hline
\rule[-0.7em]{0pt}{2.0em}Data set& $U\dmrm{corr}$& ($N$)& $B\dmrm{corr}$& ($N$)& $V\dmrm{corr}$& ($N$)\\
\hline
\rule{0pt}{1.2em}Burkhead& $-$0.12& (4)& $-$0.02& (6)& $+$0.05& (4)\\
Mt Lemmon& $-$0.03& (4)& $-$0.04& (4)& $-$0.02& (4)\\
Turku& $+$0.00& (2)& $+$0.04& (2)& $+$0.07& (2)\\
Las Campanas& $+$0.02& (2)& $+$0.00& (0)& $+$0.07& (7)\\
\hline
\end{tabular}
\end{table}

Since the Geneva photometry set of observations is the most complete and the most accurate, we chose to rescale the other sets of magnitudes according to the Geneva photometric system.
Similarly to Belokon' (\cite{B91}), we adjusted each set of magnitudes by comparing pairs of nearly simultaneous observations ($\Delta\,t\leq 5$\,days).
We first determined with 35 to 40 such pairs of observations taken mainly during 1986, that the Crimean magnitudes are consistent (average deviation smaller than 0.01 magnitude) with the Geneva photometry flux densities if their zero-magnitude fluxes are 1700\,Jy (U), 3900\,Jy (B) and 3600\,Jy (V).
These two sets of observations being consistently linked by these values, we corrected the other sets of magnitudes according to this combined reference data set.
The magnitude corrections we applied are given in Table~\ref{UBVcorr}, together with the number $N$ of observation pairs we used.
We then converted the Crimean original magnitudes and the other corrected magnitudes to flux densities using the zero-magnitude fluxes given above.
Finally, we completed the UBV light curves by including a few isolated flux densities reported by Landau et al. (\cite{LJE83}, \cite{LGJ86}) and Sadun (\cite{S85}).

\begin{figure*}[tb]
\includegraphics[width=12cm]{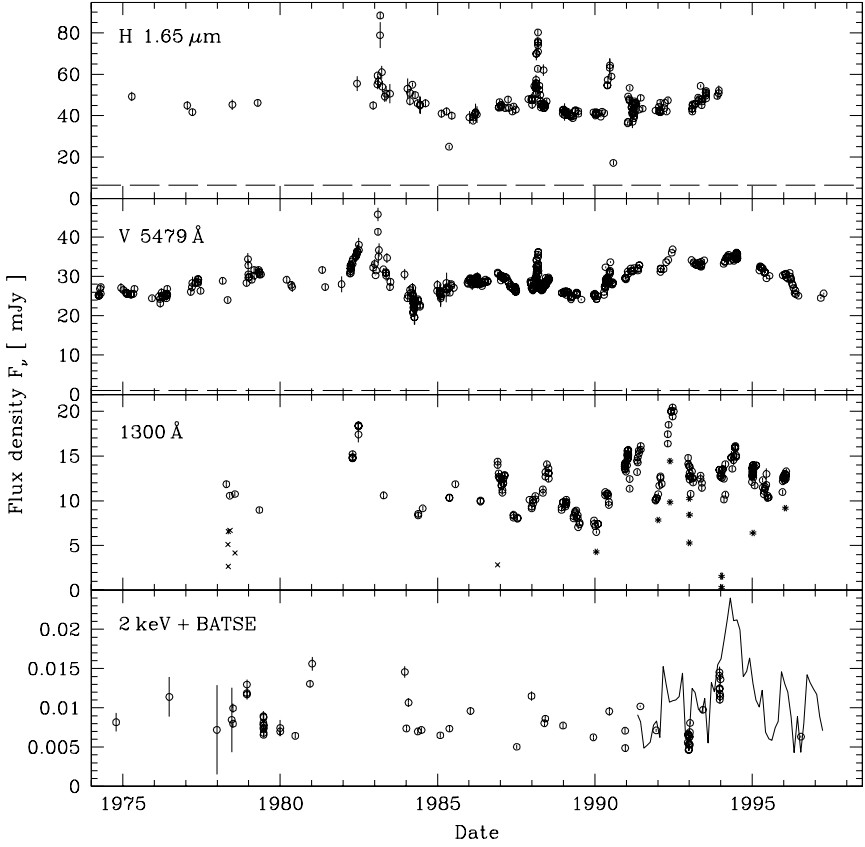}
\hfill
\parbox[b]{55mm}{
\caption{Four characteristic light curves of the infrared-to-X-ray behaviour of \3c\ from 1974 to 1998.
Three flares are clearly visible in the H and V band light curves in 1983, 1988 and 1990.
The dashed line is the contribution from the host galaxy (see Sects.~\ref{IR} and \ref{O}).
In the 1300\,\AA\ ultraviolet light curve, we indicated by crosses the six small aperture IUE observations and by stars the 11 other dubious spectra (see Sect.~\ref{UV}).
We added to the 2\,keV light curve the BATSE 20--350\,keV light curve (solid line), which was rebinned into 0.1 year bins and was extrapolated to 2\,keV assuming a power law of spectral index $\alpha$\,=\,0.6}
\label{lcirouvx}
}
\end{figure*}

The obtained V band light curve is shown in Fig.~\ref{lcirouvx}.
The contribution from the host galaxy in the V band is only 1.0\,mJy.
This value was obtained from the host galaxy's V magnitude of 16.4 (Bahcall et al. \cite{BKS97}) by using the V band zero-magnitude flux of 3647\,Jy given by Elvis et al. (\cite{EWM94}).
The relative contribution of the stars in the host galaxy is only about 3--4\,\% in the V band, which is about four times less than in the H band (see Sect.~\ref{IR} and Fig.~\ref{avspect}).

The database contains also observations from the literature in the R (7000\,\AA) and I (9000\,\AA) filters, which are not included in the Geneva photometry.
The R and I magnitudes added to the database are from Cutri et al. (\cite{CWR85}), Elvis et al. (\cite{EWM94}), Hamuy \& Maza (\cite{HM87}) (24$\arcsec$ aperture), Impey et al. (\cite{IMT89}), Moles et al. (\cite{MGM86}), O'Dell et al. (\cite{OPS78}), Smith et al. (\cite{SBE87}), Takalo et al. (\cite{TSN92}) and Valtaoja et al. (\cite{VVS91}).
All these magnitudes were converted into flux densities using the zero-magnitude fluxes given by Elvis et al. (\cite{EWM94}), which are 2791\,Jy (R) and 1871\,Jy (I).
We also added the R and I band flux densities of \3c\ reported by Landau et al. (\cite{LJE83}, \cite{LGJ86}), Lichti et al. (\cite{LBC95}), Sadun (\cite{S85}) and von Montigny et al. (\cite{VAA97}).

\section{\label{UV}Ultraviolet observations}

\3c\ was already observed by the International Ultraviolet Explorer (IUE) satellite in the first month of the mission.
As soon as the existence of important variability was established (Courvoisier \& Ulrich \cite{CU85}), a long-term monitoring campaign was launched, which started in 1986 and went on up to the end of the mission in 1996.
The spectra have been taken in the low-dispersion mode, and the usual observation rate was once every 2--3 weeks during two annual observation periods of about 3 months.

During the whole IUE mission, and taking into account only the low-dispersion observations, we have collected 256 short wavelength (SWP: 1150--1980\,\AA) spectra and 212 long wavelength (LWP or LWR: 1850--3350\,\AA) spectra.
The spectra presented here are the IUE newly extracted spectra (INES) that were taken from the INES access catalogue publicly available on the WWW at: http://ines.vilspa.esa.es/ines/.
This site contains also an important documentation including the quality flag description.
The INES flux extraction algorithm (Rodr\'{\i}guez-Pascual et al. \cite{RSW98}) was built to correct some problems found in the spectra of the IUE final archive (IUEFA) reduced with the NEWSIPS software.
The comparison of IUE flux extraction by INES and NEWSIPS shows that the INES spectra are generally more reliable than the NEWSIPS spectra in difficult conditions (Schartel \& Skillen \cite{SS98}).

\begin{figure}[tb]
\includegraphics[width=\hsize]{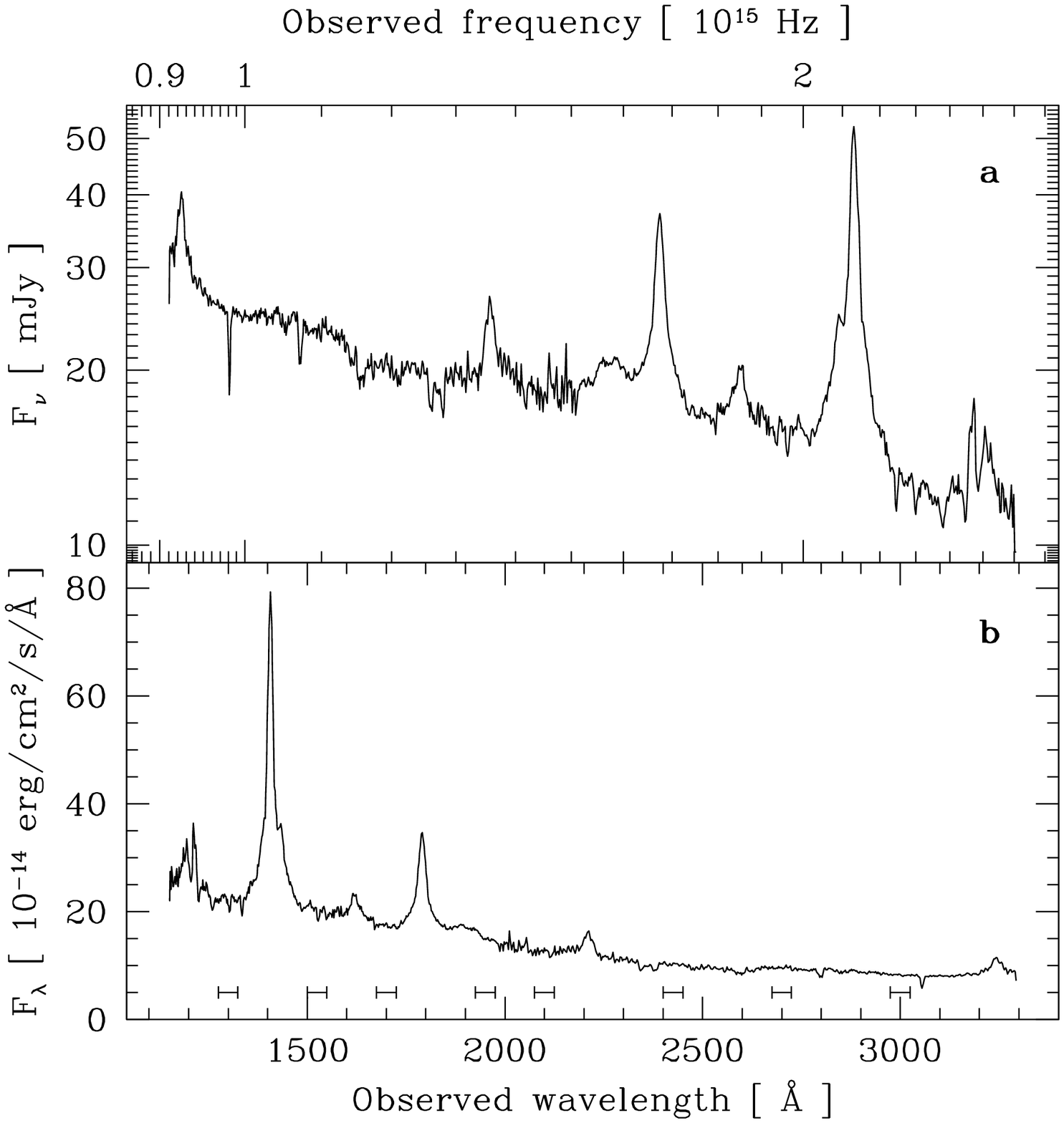}
\caption{\textbf{a and b.} The average IUE spectrum of \3c\ in the range \Lobs\,1150--3295\,\AA. \textbf{a} the logarithmic $F_{\nu}$ vs. $\nu$ representation. \textbf{b} the $F_{\lambda}$ vs. $\lambda$ representation with the continuum bands from which the UV light curves were extracted}
\label{iue}
\end{figure}

The ultraviolet (UV) continuum light curves (see Table~\ref{tabirouv}) were extracted from eight 50\,\AA\ continuum bands centered at \Lobs\,1300\,\AA, 1525\,\AA, 1700\,\AA, 1950\,\AA, 2100\,\AA, 2425\,\AA, 2700\,\AA\ and 3000\,\AA\ (see Fig.~\ref{iue}).
We considered only the wavelength bins with a quality flag of zero.
The spectra having no such bin in a continuum band are therefore not included in the corresponding light curve (e.g. SWP01365LL).

The flux density $F_{\nu}$ and its uncertainty $\Delta\,F_{\nu}$ in a 50\,\AA\ band were defined as
\begin{equation}
\label{erriue}
F_{\nu}=\frac{1}{n}\sum_{i=1}^{n}F_{\nu_i}\quad\mbox{and}\quad\Delta\,F_{\nu}=\frac{x}{n}\sqrt{\sum_{i=1}^{n}(\Delta\,F_{\nu_i})^2}
\end{equation}
where $F_{\nu_i}$ and $\Delta\,F_{\nu_i}$ are respectively the INES flux density and its uncertainty at the frequency $\nu_i$, and $x$ is a correction factor.
The factor $x$ is introduced to ensure that the average uncertainty $\overline{\Delta\,F_{\nu}}$ corresponds to the experimental value derived from all pairs of observations taken within one day.
The obtained values of $x$ reported in Table~\ref{tabxiue} are not simply equal to one, because the bins used to calculate the fluxes are not independent from each other due to the oversampling of the IUE spectra.
However, by calculating $x$ for a single INES bin, we obtain on average over the eight values given in Table~\ref{tabxiue} a value of 0.97\,$\pm$\,0.36, which shows that the INES uncertainties for a single bin are in good agreement with the experimental values derived from consecutive spectra.
Our estimation of the uncertainties gives average relative values, $\overline{\Delta\,F_{\nu}}/\overline{F_{\nu}}$, around 2--3\,\% in all eight 50\,\AA\ bands.

\begin{table}[tb]
\caption{Values of the correction factor $x$ in Eq.~(\ref{erriue}) determined for a 50\,\AA\ band and a single INES bin at the observed wavelength \Lobs}
\label{tabxiue}
\begin{tabular}{@{}ccc|ccc@{}}
\hline
\rule[-0.7em]{0pt}{2.0em}\Lobs& $x$ (50\,\AA)& $x$ (1\,bin)& \Lobs& $x$ (50\,\AA)& $x$ (1\,bin)\\
\hline
\rule{0pt}{1.2em}1300\,\AA& 1.9& 0.9& 2100\,\AA& 0.7& 0.5\\
1525\,\AA& 2.0& 0.9& 2425\,\AA& 1.1& 0.6\\
1700\,\AA& 3.0& 1.6& 2700\,\AA& 1.4& 0.8\\
1950\,\AA& 4.6& 1.4& 3000\,\AA& 2.0& 1.0\\
\hline
\end{tabular}
\end{table}

In Fig.~\ref{lcirouvx} we show the 1300\,\AA\ UV light curve extracted from the IUE spectra at \Lobs\,1275--1325\,\AA.
It can be seen that the six SWP small aperture spectra (namely, SWP01492LS, SWP01498LS, SWP01509LS, SWP01655LS, SWP02100LS and SWP29775LS) all lie well below the light curve drawn only with the large aperture observations.
The same occurs with the two LWR small aperture spectra (LWR01447LS and LWR04470LS), clearly illustrating that, even for point sources, the small aperture of IUE gives unreliable fluxes due to the relatively large size of the IUE point spread function compared to the size of the small aperture.

Other points lie well outside the general UV light curves.
In three spectra (SWP49812LL, SWP49813LL and LWP27215LL) no significant flux is detected, because of a technical problem.
Other spectra show a characteristic \3c\ spectrum, but with a normalization very different from the typical flux at this period.
Most of these spectra have a problem which is reported in the INES header of the spectrum or at least in the SILO file.
SWP38038LL and SWP53305LL are out of aperture.
SWP43550LL, SWP46633LL, SWP46647LL, SWP46650LL, LWP22191LL, LWP24638LL and LWP24657LL had no tracking during exposure and thus drifted out.
LWP27045LL is contaminated by the solar spectrum and LWP24741LL had no guiding and is most probably also contaminated by scattered solar spectrum.

Three other SWP spectra and one LWP spectrum (LWP21967LL) are dubious, although no problem has been reported.
The SWP spectra SWP44732LL, SWP44733LL and SWP56622LL have a \Lya\ emission-line flux which is respectively only 82\,\%, 60\,\% and 73\,\% that of a comparison spectrum obtained at nearby epoch.
It suggests that there is a normalization problem with these spectra, and not an unusually fast variability, since the \Lya\ line was observed to remain constant within $\pm$\,8\,\% of the average line flux from 1978 to 1992 (Ulrich et al. \cite{UCW93}).
This argument cannot be applied to the LWP spectrum LWP21967LL, but it was found to have a flux about 20\,\% lower than that of the following spectrum obtained only three hours later.

All the spectra discussed above are flagged in the UV light curves by a non-zero value of ``Flag''.
Small aperture spectra are flagged with a value of $-$2, whereas other spectra with problems or found to be dubious are flagged with a value of $-$1.
The average IUE spectrum of \3c\ shown in Fig.~\ref{iue} was constructed by excluding these spectra, but without taking into account the INES quality flag of each wavelength bin.

\section{\label{X}X-ray and $\gamma$-ray observations}

The database contains X-ray light curves at 0.1, 0.2, 0.5, 1, 2, 5, 10, 20, 50, 100, 200 and 500\,keV and $\gamma$-ray light curves at 1\,MeV, 3\,MeV, 100\,MeV, 300\,MeV and 1\,GeV (see Table \ref{tabxg}).
These light curves were constructed with the spectral fit parameters given in the literature.
All X-ray observations of \3c\ until 1990 are summarized in Malaguti et al. (\cite{MBC94}).
We completed this list to the best of our knowledge with the more recent X-ray and $\gamma$-ray observations of \3c\ found in the literature.
The parameters we used were sometimes directly taken from Malaguti et al. (\cite{MBC94}), but we often tried to find additional information in the original publications.
We usually preferred to use the parameters from fits in which the absorbing hydrogen column density \nh\ was fixed to its galactic value ($\sim$\,1.8\,10$^{20}$\,atoms\,cm$^{-2}$), rather than left as a free parameter of the fit.

Apart from Malaguti et al. (\cite{MBC94}), we used the spectral fit parameters from the following references.
The European X-ray Observatory Satellite (EXOSAT) observations and most Ginga observations are from Turner et al. (\cite{TWC90}).
The Roentgen Observatory Satellite (ROSAT) observations are from Staubert et al. (\cite{SFC92}), B\"uhler et al. (\cite{BCS95}) and Leach et al. (\cite{LMP95}).
The CGRO observations are from Johnson et al. (\cite{JDK95}), Lichti et al. (\cite{LBC95}), McNaron-Brown et al. (\cite{MJJ95}) and von Montigny et al. (\cite{VAA97}).
The Advanced Satellite for Cosmology and Astrophysics (ASCA) observations are from Cappi et al. (\cite{CMO98}), the SIGMA observations are from Churazov et al. (\cite{CGF94}), the Extreme Ultraviolet Explorer (EUVE) observations are from Ramos et al. (\cite{RKF97}) and the ``Satellite per Astronomia X'' (SAX) observation is from Grandi et al. (\cite{GGM97}).

\begin{table}[tb]
\caption{The X- and $\gamma$-ray light curves of \3c\ in the database. The parameters are as in Table \ref{tabrmm}, with $\overline{F_{\nu}}$ and $\sigma_{\nu}$ expressed in Jy}
\label{tabxg}
\begin{tabular}{@{}l@{~~~}c@{~~~}rcl@{~~~}l@{}}
\hline
\rule[-0.7em]{0pt}{2.0em}Light curve& Epoch& \multicolumn{1}{c}{$N$}& $\overline{\nu}$& \multicolumn{1}{c}{$\overline{F_{\nu}}$}& \multicolumn{1}{c}{$\sigma_{\nu}$}\\
\hline
\rule{0pt}{1.2em}0.1\,keV& 1979--95& 22& 2.42\,10$^{16}$& 2.48\,10$^{-4}$& 1.0\,10$^{-4}$\\
0.2\,keV& 1979--93& 20& 4.84\,10$^{16}$& 9.47\,10$^{-5}$& 3.3\,10$^{-5}$\\
0.5\,keV& 1969--93& 32& 1.21\,10$^{17}$& 2.90\,10$^{-5}$& 7.6\,10$^{-6}$\\
1\,keV& 1969--96& 44& 2.42\,10$^{17}$& 1.44\,10$^{-5}$& 3.8\,10$^{-6}$\\
2\,keV& 1969--96& 68& 4.84\,10$^{17}$& 8.67\,10$^{-6}$& 2.9\,10$^{-6}$\\
5\,keV& 1969--96& 46& 1.21\,10$^{18}$& 5.86\,10$^{-6}$& 1.9\,10$^{-6}$\\
10\,keV& 1969--96& 44& 2.42\,10$^{18}$& 4.15\,10$^{-6}$& 1.4\,10$^{-6}$\\
20\,keV& 1976--96& 21& 4.84\,10$^{18}$& 3.26\,10$^{-6}$& 2.0\,10$^{-6}$\\
50\,keV& 1977--96& 36& 1.21\,10$^{19}$& 1.62\,10$^{-6}$& 8.2\,10$^{-7}$\\
BATSE& 1991--97& 2133& 1.77\,10$^{19}$& 1.30\,10$^{-6}$& 1.3\,10$^{-6}$\\
100\,keV& 1978--96& 34& 2.42\,10$^{19}$& 1.12\,10$^{-6}$& 6.4\,10$^{-7}$\\
200\,keV& 1978--96& 29& 4.84\,10$^{19}$& 8.17\,10$^{-7}$& 5.9\,10$^{-7}$\\
500\,keV& 1990--95& 14& 1.21\,10$^{20}$& 2.88\,10$^{-7}$& 1.7\,10$^{-7}$\\
1\,MeV& 1990--95& 15& 2.42\,10$^{20}$& 1.80\,10$^{-7}$& 1.1\,10$^{-7}$\\
3\,MeV& 1991--91& 1& 7.25\,10$^{20}$& 5.63\,10$^{-8}$&\multicolumn{1}{c}{---}\\
100\,MeV& 1976--93& 5& 2.42\,10$^{22}$& 4.76\,10$^{-10}$& 2.4\,10$^{-10}$\\
300\,MeV& 1976--93& 5& 7.25\,10$^{22}$& 8.77\,10$^{-11}$& 4.0\,10$^{-11}$\\
1\,GeV& 1991--93& 4& 2.42\,10$^{23}$& 1.36\,10$^{-11}$& 8.8\,10$^{-12}$\\
\hline
\end{tabular}
\end{table}

From the spectral fit parameters we constructed the light curves as follows.
For each observation, we first derived the mean photon energy $E\dmrm{mean}$ in the energy range $E_1$--$E_2$ of the fit knowing the spectral index $\alpha$.
We then calculated the flux normalization $F\dmrm{norm}$ at $E\dmrm{mean}$ and its uncertainty $\Delta\,F\dmrm{norm}$.
Finally, we obtained the flux densities $F_E$ at the photon energies $E$ of the light curves between $E_1$ and $E_2$ by the relation
\begin{equation}
F_E=F\dmrm{norm}(E/E\dmrm{mean})^{-\alpha} \quad \mbox{where} \quad E\in[\,E_1\,;\,E_2\,]\,.
\end{equation}
The corresponding flux uncertainties $\Delta\,F_E$ were derived from both $\Delta\,F\dmrm{norm}$ and $\Delta\,\alpha$ according to the equation
\begin{eqnarray*}
\Delta\,F_E&=&\sqrt{\left(\frac{\partial F_E}{\partial F\dmrm{norm}}\Delta\,F\dmrm{norm}\right)^2+\left(\frac{\partial F_E}{\partial \alpha}\Delta\,\alpha\right)^2}\\
&=&F_E\sqrt{(\Delta\,F\dmrm{norm}/F\dmrm{norm})^2+(\ln{(E/E\dmrm{mean})\Delta\,\alpha})^2}\,.
\end{eqnarray*}
As far as possible, we always converted the uncertainties from the literature into 1-$\sigma$ uncertainties (68\,\% confidence level).
$\Delta\,F\dmrm{norm}$ and $\Delta\,\alpha$ were respectively assumed to be 5\,\% of $F\dmrm{norm}$ and 0.5, if the corresponding uncertainties were not found in the reference.
If both upper and lower uncertainties are given by the authors, we fixed the uncertainty to the smaller value.

We also included in the database the Burst and Transient Source Experiment (BATSE) light curve of \3c\ derived from Earth occultation data, which are made public on the WWW by the Compton Observatory Science Support Center (COSSC).
The BATSE light curve was derived from the daily photon fluxes in the range 20--350\,keV assuming a photon spectral index $\Gamma$ of 1.7.
Its frequency corresponds to the mean photon energy in the range 20--350\,keV with $\Gamma$\,=\,1.7.
We used this value of $\Gamma$, because it is the value assumed when the occultation data are reduced.
However, a $\Gamma$ of 1.6 would better correspond to the 1--200\,keV SAX observation reported by Grandi et al. (\cite{GGM97}).
Since the hard X-ray spectral index of \3c\ does not vary much, we show in Fig.~\ref{lcirouvx} the general shape of the BATSE light curve extrapolated to 2\,keV with a constant $\Gamma$ of 1.6.
This light curve is in fairly good agreement with the contemporary 2\,keV observations.

\begin{figure*}[tb]
\includegraphics[width=12cm]{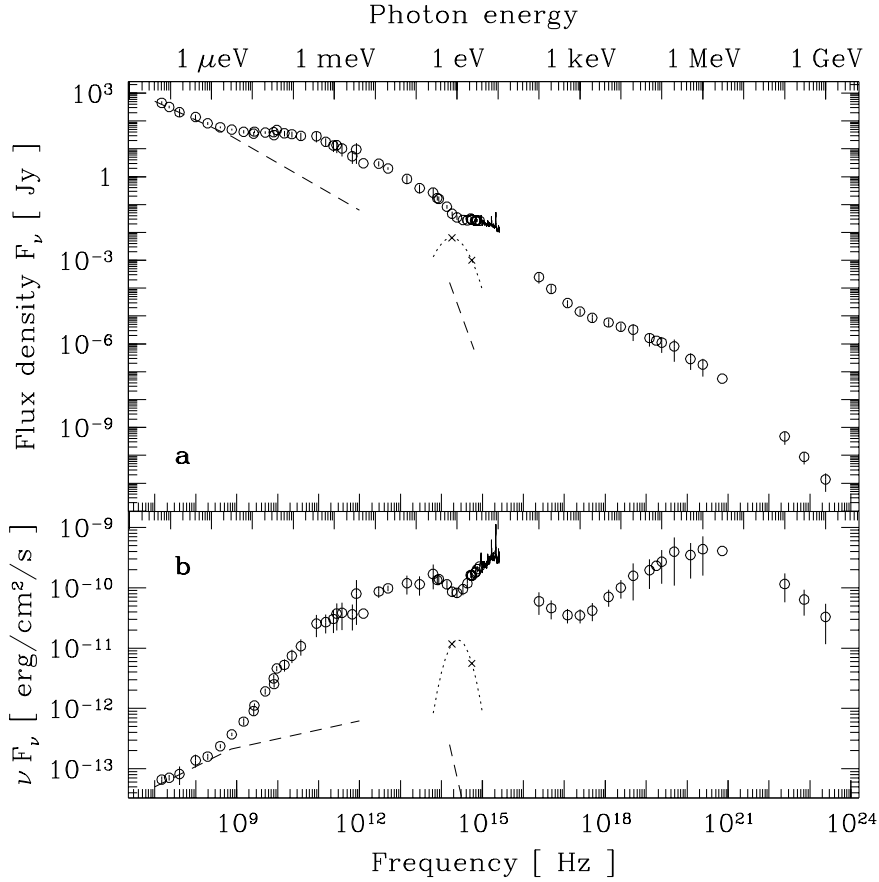}
\hfill
\parbox[b]{55mm}{
\caption{\textbf{a and b.} Average spectrum of \3c\ constructed with the parameters $\overline{\nu}$, $\overline{F_{\nu}}$ and $\sigma_{\nu}$ given in Tables \ref{tabrmm}, \ref{tabirouv} and \ref{tabxg} (points) and the average IUE spectrum (solid line).
\textbf{a} the $F_{\nu}$ representation; \textbf{b} the $\nu\,F_{\nu}$ representation.
The dashed line is the contribution from the jet (3C\,273A) (see Sect.~\ref{R}).
The contribution from the host galaxy is suggested by a parabola in $\nu\,F_{\nu}$ (dotted line), which is normalized by the H and V band contributions (crosses) (see Sects.~\ref{IR} and \ref{O}).
The relatively small error-bars show that the variability of \3c\ does not basically change its overall spectral shape}
\label{avspect}
}
\end{figure*}

\section{\label{database}Description of the database}

The overall database is illustrated in Fig.~\ref{allobs} by a time versus frequency representation of all observations currently in the database and in Fig.~\ref{avspect} by the average spectrum of \3c.
The observations presented here are publicly available on the WWW at the address: http://obswww.unige.ch/3c273/.
Special care was taken to have a homogeneous and clear database, which can be easily understood and used.
The user should however be aware that the presented light curves often contain observations at different (but close) wavelengths.
To facilitate the selection of observations within the light curves, each individual measurement is always characterized by the epoch of observation, the frequency, the wavelength, the flux, the flux uncertainty, the observatory that made the observation and the literature reference of previously published observations.

The epoch of observation is given with an accuracy of up to one hour in universal time (UT).
It is both expressed by the year (yyyy), the month (mm), the day (dd) and the hour (hh) coded in the number ``yyyymmdd.hh'' and in fractional years of same duration (365.25\,days).
The fractional year ``Date'' is given with an accuracy of 0.0001\,year, which is about one hour (1\,day $\simeq$ 0.0027\,year).
It was always derived from the Julian date (JD) by the following relation:
\begin{equation}
\mbox{Date}=2000.0+(\mbox{JD}-2\,451\,544.5)/365.25
\end{equation}
If an observation lasted more than one day, the epoch is set to the middle of the observation period.
If ``yyyymmdd'' finishes by two or four zeros, it means respectively that the day or the month of the observation was not known to us.
In this case, the value of ``Date'' was calculated for the middle of the month or of the year respectively.
For a few radio catalogue data where we could not find any indication about the date of observation, we fixed the epoch to the year of publication, since it is a way to identify them.

We chose to give both the observed frequency and the observed wavelength at which the flux measurement was made.
The frequency ``Nu'' ($\nu$) is always expressed in Hz and the wavelength ``Lambda'' ($\lambda$) is always expressed in \AA.
The flux density ``Fnu'' ($F_{\nu}$) and its uncertainty ``Delfnu'' ($\Delta\,F_{\nu}$) are always expressed in Jansky (1\,Jy\,=\,10$^{-23}$\,\ergcmsHz\,=\,10$^{-26}$\,W\,m$^{-2}$\,Hz$^{-1}$).
Upper limits are not considered in this database.
A zero value for ``Delfnu'' means that the relevant uncertainty is not known to us.
The observatory, the telescope site, or the satellite that made the observation is indicated by a name or an abbreviation in the column labeled ``Obs''.
For published data, the reference ``Ref'' is given in an abbreviated form (initial letters of the three first authors' surnames and the year (yy) of publication).
Instead of the ``Ref'' column, the UV light curve files contain two other columns ``Spectrum'' and ``Flag'' containing respectively the INES spectrum identification and our general quality flag defined in Sect.~\ref{UV}.

The average spectrum shown in Fig.~\ref{iue} and all individual IUE spectra are given in one file per observation in ASCII format.
The files are simply the ASCII equivalent of the FITS files from the INES catalogue and the original headers are also provided.
The date of observation and our general quality flag can be found in the UV light curves.

The original X-ray spectral fit parameters, which we used for constructing the light curves are also provided in a file.
This file contains for each observation the epoch of its start and sometimes of its end, the energy range $E_1$--$E_2$ of the fit, the photon spectral index $\Gamma$\,=\,$\alpha$\,$+$\,1, the flux normalization at 1\,keV, $F\dmrm{1keV}$, in units of 10$^{-2}\,\phcmskeV$ (which corresponds to 6.6261 $\mu$Jy) and the associated uncertainties $\Delta\,\Gamma$ and $\Delta\,F\dmrm{1keV}$.
We also indicate the confidence level of the quoted uncertainties in $\sigma$ (1\,$\sigma$\,=\,68\,\%, 1.6\,$\sigma$\,=\,90\,\%, 2\,$\sigma$\,=\,95\,\%), whether the fit was performed with a free value of the hydrogen column density \nh\ or not, the instrument that made the observation and the reference.

\section{\label{conclu}Conclusion}

This paper is the witness of the enormous observing effort made all over the world to monitor \3c\ at every possible wavelength during more than 30 years.
It is to the best of our knowledge the first attempt of making an archive grouping together most available observations of a single AGN through the entire electro-magnetic spectrum.
Such a database, apart from making data publicly available, ensures that early observations sometimes only published in graphical form are not completely lost after a few decades.

The aim of the database presented here is to make the analysis of multi-wavelength variability possible for a large community of astronomers.
We are convinced that such studies are of great help to the understanding of the physical processes at work in \3c.
This paper might also stimulate the creation of similar archives for other well observed AGN (Seyfert galaxies or blazars), which would enable the direct comparison of variability between different object classes and hence constrain the unification models of AGN.

\begin{acknowledgements}
We are very grateful to all observers and collaborators, who participated in the Geneva photometric monitoring of \3c.
We also thank D.C. Backer and R. Barvainis for providing us the 89\,GHz observations from Hat Creek and the 87\,GHz FCRAO observations, respectively.
M. Aller and H. Aller would like to acknowledge partial support by the National Science Foundation (NSF) from grant AST 94-21\,979 and preceding grants.
The BIMA array is operated by the Berkeley-Illinois-Maryland Association under funding from the NSF under grant AST 93-20\,238.
The GBI was operated by the National Radio Astronomy Observatory under contract to the US Naval Observatory and the Naval Research Laboratory (NRL) during the time period of these observations.
Radio astronomy at the NRL is supported by the Office of Naval Research.
\end{acknowledgements}

\end{document}